\documentclass{osa-article}

\journal{osajournal}

\usepackage{enumitem}
\usepackage[utf8]{inputenc}

\articletype{Research Article}

\begin{document}


\title{Geometry-dependent two-photon absorption followed by free-carrier absorption in AlGaAs waveguides}

\author{Daniel H. G. Espinosa,\authormark{1,\dag} Stephen R. Harrigan,\authormark{2,\dag} Kashif M. Awan,\authormark{3} Payman Rasekh,\authormark{1} and Ksenia Dolgaleva\authormark{1,2,*}}

\address{\authormark{1}School of Electrical Engineering and Computer Science, University of Ottawa, Ottawa, Ontario K1N 6N5, Canada\\
\authormark{2}Department of Physics, University of Ottawa, Ottawa, Ontario K1N 6N5, Canada\\
\authormark{3}Stewart Blusson Quantum Matter Institute, Vancouver, BC V6T 1Z4, Canada\\
\authormark{\dag}These authors contributed equally to this work.}

\email{\authormark{*}ksenia.dolgaleva@uottawa.ca} 



\begin{abstract}
Nonlinear absorption can limit the efficiency of nonlinear optical devices. However, it can also be exploited for optical limiting or switching applications. Thus, characterization of nonlinear absorption in photonic devices is imperative for designing useful devices. This work uses the nonlinear transmittance technique to measure the two-photon absorption coefficients ($\alpha_2$) of AlGaAs waveguides in the strip-loaded, nanowire, and half-core geometries in the wavelength range from $1480$ to $1560~\text{nm}$. The highest $\alpha_2$ values of $2.4$, $2.3$, and $1.1~\text{cm}/\text{GW}$ were measured at $1480~\text{nm}$ for a $0.8$-nm-wide half-core, $0.6$-nm-wide nanowire, and $0.9$-nm-wide strip-loaded waveguides, respectively, with $\alpha_2$ decreasing with increasing wavelength. The free-carrier absorption cross-section was also estimated from the nonlinear transmittance data to be around $2.2\times10^{-16}~\text{cm}^2$ for all three geometries. Our results contribute to a better understanding of the nonlinear absorption in heterostructure waveguides of different cross-sectional geometries. We discuss how the electric field distribution in the different layers of a heterostructure can lead to geometry-dependent effective two-photon absorption coefficients. More specifically, we pinpoint the third-order nonlinear confinement factor as a design parameter to estimate the strength of the effective nonlinear absorption, in addition to tailoring the bandgap energy by varying the material composition.
\end{abstract}

\section{Introduction}

Aluminum gallium arsenide (AlGaAs) is an established material platform among III-V semiconductors considered for nonlinear photonics applications. The nonlinear optical properties of waveguides made of AlGaAs heterostructures on GaAs substrates, and, more recently, on SiO$_{2}$ and sapphire substrates, have been exploited to demonstrate efficient four-wave mixing~\cite{espinosa2021tunable,pu2018ultra,dolgaleva:2015}, orthogonal four-wave mixing~\cite{johnson2019orthogonal}, wavelength conversion~\cite{kaminski2019characterization,da2017characterization}, frequency comb generation~\cite{hu2018single}, sum- and difference-frequency generation~\cite{savanier2011nearly} and second-harmonic generation~\cite{may2019second,duchesne2011second}. Additionally, the nearly perfect lattice matching between GaAs and AlAs allows for arbitrary tuning of the aluminium fraction $x$ in an Al$_{x}$Ga$_{1-x}$As alloy. By tailoring the aluminium fraction, which impacts the material properties such as the bandgap energy, devices can be fabricated with engineered dispersion and tunable linear and nonlinear absorption and refraction~\cite{dolgaleva:2015,gehrsitz2000refractive,stegeman1994algaas,bosio1988direct,adachi1988optical}.

Nonlinear optical interactions are often accompanied by nonlinear absorption (NLA), which can be viewed as either a limit in some cases or produce usable effects in others. For instance, in wave-mixing or harmonic generation processes, the incident photons absorbed by the device's material through nonlinear absorption mechanisms do not contribute to the generation of new frequency components. In these cases, the efficiency of the nonlinear optical process is reduced because of the NLA~\cite{espinosa2021tunable}. For this reason, it is important to perform a thorough nonlinear absorption (NLA) characterization of new materials and devices to understand their practical limit. On the other hand, the usefulness of nonlinear absorption can be exemplified by optical limiting or all-optical switching devices. Optical limiting devices protect downstream optical and optoelectronic components, such as detectors and sensors, by allowing the transmission of the input light beam at low-irradiance levels and limiting the transmission of damaging levels.~\cite{van1988optical}. All-optical switching devices based on microring resonator or photonic crystal (PhC) cavities exploit the change in the refractive index caused by the 2PA-generated free carriers~\cite{nozaki2010sub,van:2002}. The resonant wavelength of the ring or PhC cavity is modified, during the lifetime of the 2PA-generated free carriers. Thus, a low-power probe beam, coupled to the resonant wavelength, has its transmission controlled by a 2PA-inducing pump beam. Hence, investigating the influence of the waveguide geometry on the effective two-photon absorption coefficient is crucial for predicting and designing useful devices.

We have recently reported on the experimental study of four-wave mixing (FWM) in AlGaAs waveguides of three different geometries: strip-loaded, nanowire, and half-core [see Fig.~\ref{fig:waveguide-designs} (c)--(e), showing the cross-section of each geometry]~\cite{espinosa2021tunable}. In the strip-loaded waveguide, the light propagates in the slab region below the ridge defined in the upper cladding, whereas in the nanowire, the guiding layer is etched through. The nanowire allows the realization of narrower waveguides for stronger light confinement, while the strip-loaded waveguide exhibits the lowest propagation loss. The half-core waveguide geometry represents a compromise between the other two, featuring stronger light confinement than the strip-loaded design but lower propagation loss than the nanowire design. A comparative analysis of FWM indicated that the NLA of a pump beam with a pulse width of $3~\text{ps}$ impacts the FWM efficiency of each geometry differently, mainly for the wavelengths longer than $1500~\text{nm}$. This result motivated further investigations to better understand NLA processes in these geometries~\cite{espinosa2021tunable}.

Nonlinear absorption in AlGaAs strip-loaded waveguides with Al$_{0.18}$Ga$_{0.82}$As as the guiding layer composition has been studied for waveguide widths ranging from $5$ to $6~\mu\text{m}$~\cite{aitchison:1997,kang1994limitation,villeneuve:1993}. The 2PA coefficient in a $1.5$-$\mu$m-wide nanowire with Al$_{0.13}$Ga$_{0.87}$As as the guiding layer composition has also been measured~\cite{wathen2014efficient}. However, there are no reports on the influence of the nonlinear absorption on the nonlinear optical performance of AlGaAs nanowires with the guiding layer composition Al$_{0.18}$Ga$_{0.82}$As, and no reports on measuring 2PA in half-core waveguides, representing a recently proposed geometry~\cite{espinosa2021tunable,awan2015aluminium}. Moreover, systematic comparison of 2PA in waveguides of different geometries having the same guiding layer composition has not been performed.


Advances in nanofabrication, such as electron-beam lithography and plasma etching, made possible the realization of devices as narrow as hundreds of nanometers in width~\cite{dolgaleva:2015}. To facilitate coupling light into such devices, it is typical to use wider coupling waveguides and to taper them down to the width of the narrower part, as indicated in Fig.~\ref{fig:waveguide-designs}~(a). Using tapers, however, complicates the analysis of the nonlinear optical performance of such devices. The coupling part itself may present nonlinear effects, making the evaluation of the irradiance in the narrower parts more challenging. Furthermore, while there are reports on the presence of 2PA followed by free-carrier absorption (FCA) as a NLA mechanism in GaAs/AlGaAs waveguides~\cite{villeneuve1994nonlinear} and GaAs/AlGaAs multiple-quantum-well waveguides~\cite{laughton1992intuitive,laughton1992time}, FCA has not been quantified as a loss mechanism in the NLA measurements of AlGaAs waveguides~\cite{aitchison:1997,kang1994limitation,villeneuve:1993,wathen2014efficient}. Proper separation of these effects is crucial for a proper extraction of the 2PA coefficient from the NLA data.

In this work, we measure the 2PA coefficient in the wavelength range of $1480~\text{nm}$ to $1560~\text{nm}$, falling in the optical communications C-band, of AlGaAs waveguides of strip-loaded, nanowire, and half-core geometries, which all have Al$_{0.18}$Ga$_{0.82}$As as the guiding layer composition. This work represents the first comparative study of the influence of the NLA on various waveguide geometries under identical experimental conditions, and the first NLA measurement performed in a half-core waveguide. The data analysis includes the effect of FCA with an estimation of the FCA cross-section. Our method also addresses the effect of nonlinear absorption in the coupling parts of the waveguides to properly deduce the 2PA coefficient of the narrowest parts of the waveguides, which can be of interest for future measurements of 2PA in tapered devices. Finally, we discuss how the intensity distribution in the different layers of the waveguides influences the results.

The next sections of the manuscript are organized as follows. Section~\ref{sec:exp} presents the waveguide geometries and material compositions, the simulation results of the intensity distributions, the experimental methods, and the models of the nonlinear absorption used in this work. Section~\ref{sec:res} reports the 2PA coefficient and FCA cross-section results for each geometry as a function of wavelength. Section~\ref{sec:dis} discusses the effective 2PA coefficient dependence on the geometry-dependent nonlinear confinement factor, followed by the conclusions in Section~\ref{sec:conc}. Finally, the details on the linear loss measurements are presented in the appendix~\hyperref[sec:AppA]{A}.

\section{Experimental}
\label{sec:exp}

\subsection{AlGaAs waveguides}
%
\begin{figure*}[htpb]
\centerline{\includegraphics[width=18cm]{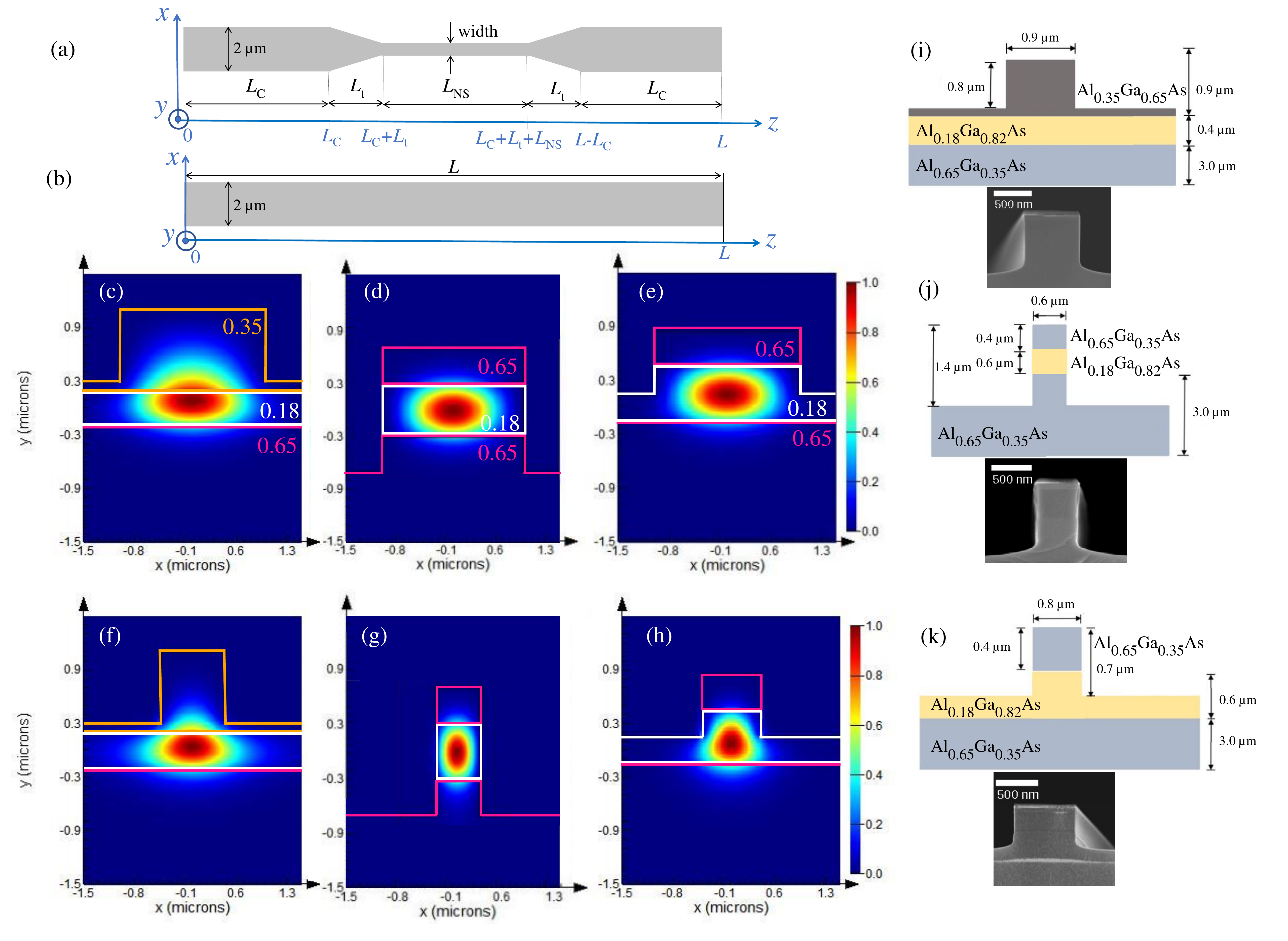}} \caption{\label{fig:waveguide-designs}(color on-line) Top-down view of the tapered (a) and reference (b) waveguides. In this work, we call the narrower part of the tapered device ``nanosectional part.'' (c)-(h) Waveguide cross-sections with the optical intensity distributions of the fundamental transverse-electric (TE) mode at $1500~\text{nm}$ in $2$-$\mu$m-wide strip-loaded (c), nanowire (d) and half-core (e) coupling waveguides, and $0.9$-$\mu$m-wide strip-loaded (f), $0.6$-$\mu$m-wide nanowire (g), and $0.8$-$\mu$m-wide half-core (h) nanosectional parts. The numbers on (c)-(h) represent the aluminum fraction $x$ of each layer. (i)-(k) Geometry, dimensions, and SEM images of the (i) strip-loaded, (j) nanowire, and (k) half-core waveguides cross-sections. $L = 5.26~\text{mm}$ and $L_\text{NS} = 1~\text{mm}$ for the strip-loaded, $L = 5.33~\text{mm}$ and $L_\text{NS} = 2~\text{mm}$ for the nanowire, and $L = 5.87~\text{mm}$ and $L_\text{NS} = 1~\text{mm}$ for the half-core waveguides. For all waveguide geometries, $L_\text{t} = 0.2~\text{mm}$ and $L_\text{C} = (L - L_\text{NS} - 2 L_\text{t})/2$.}
\end{figure*}

The waveguides of the nanowire, strip-loaded, and half-core geometries were fabricated using e-beam lithography followed by RIE/ICP dry etching. A $2$-$\mu$m-wide coupling region was used to couple the light to and from the $<1$-$\mu$m-wide (``nanosectional'') part [see Fig.~\ref{fig:waveguide-designs}~(a)]. To study the effect of the $2$-$\mu$m-wide coupling region, we prepared some straight 2-$\mu$m-wide waveguides [Fig.~\ref{fig:waveguide-designs}~(b)], called reference waveguides, on the same sample as the tapered devices. The cross-sectional intensity distributions are presented in Fig.~\ref{fig:waveguide-designs}~(c)~to~(h). The cross-sectional dimensions, the material composition, and the scanning electron microscopy (SEM) images of the waveguides facets are presented in Fig.~\ref{fig:waveguide-designs}~(i)~to~(k). The intensity distribution and the effective mode area ($A_\text{eff}$) were calculated using the finite difference eigenmode (FDE) solver of \textit{Lumerical Mode Solutions}. The FDE solver calculates the effective mode area using the equation
\begin{align}
A_\text{eff} & = \frac{ \left[ \int_{-\infty}^\infty \int_{-\infty}^\infty |E(x,y)|^2 \text{d}x \text{d}y \right]^2}{ \int_{-\infty}^\infty \int_{-\infty}^\infty |E(x,y)|^4\text{d}x \text{d}y },\label{eq:aeff}
\end{align}
which corresponds to the effective mode area calculations for the third-order nonlinear optical effects.

Other details of the waveguides design optimization, fabrication, and characterization are presented in references~\cite{awan:2018,espinosa2021tunable}.

\subsection{Experimental setup}

\begin{figure}[htbp]
\centerline{\includegraphics[width=8cm]{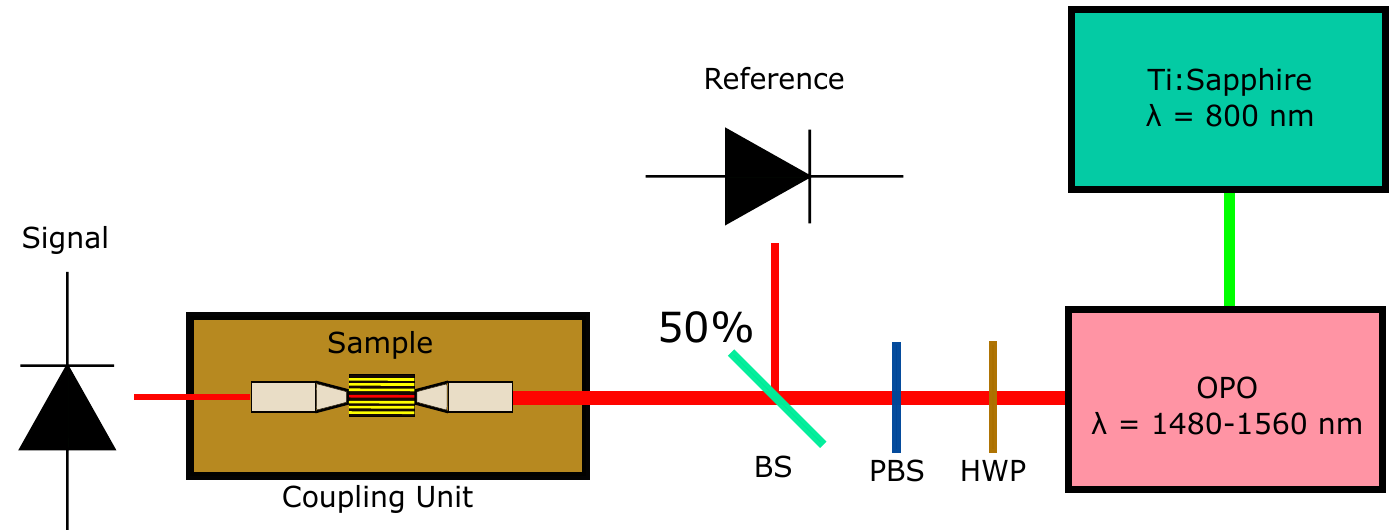}} \caption{\label{fig:exp-setup}(color on-line) Diagram of the experimental setup used to perform the nonlinear transmittance  technique. OPO - optical parametric oscillator, HWP - half-wave plate, PBS - polarizing beam splitter, BS - beam splitter.}
\end{figure}

A schematic of the experimental setup is shown in Fig.~\ref{fig:exp-setup}. We used a Ti:Sapphire laser and optical parametric oscillator (OPO), with the temporal full width at half-maximum of $\tau_\text{FWHM} = 3~\text{ps}$ and the repetition rate of $f_\text{p} = 76.6~\text{MHz}$. While the OPO wavelength can range from $1000~\text{nm}$ to about $3700~\text{nm}$, we selected the wavelengths  from $\lambda = 1480~\text{nm}$ to $\lambda = 1560~\text{nm}$, which are the range limits of the CW laser (Santec TSL-710) used to measure the linear loss coefficients ($\alpha_1$) of the reference devices. Besides, this range is the same as used to observe FWM in our devices~\cite{espinosa2021tunable}. The rotation of the half-wave plate controls the beam power while the polarizing beam splitter keeps the polarization constant. We used the objective lenses in the coupling stage to couple the light to and collect the light from the waveguides. The average power at the waveguide output ($P_{\text{out}}$) is measured as a function of the average power at the waveguide input ($P_{\text{in}}$) by the use of photodetectors and a power meter.
%

\subsection{Linear propagation loss measurement}

The linear loss coefficient $\alpha_1$, also called propagation loss coefficient, and the Fresnel coefficient [$R = 0.30(1)$] were measured in the reference waveguides by the Fabry-Perot method, using a tunable CW laser beam (model TSL-710, Santec inc.)~\cite{espinosa2021tunable,dolgaleva:2015,tittelbach1993comparison}. The linear loss coefficients of the nanosectional parts were measured using the linear region of the plot $P_{\text{out}}$ vs. $P_{\text{in}}$. The coupling efficiencies ($\eta_\text{C}$) were determined as $0.27(1)$, $0.11(1)$ and $0.20(3)$ for the strip-loaded, nanowire and half-core geometries, respectively. As the main focus of this work is the nonlinear absorption, the linear loss coefficient ($\alpha_1$) values and the related discussions are presented in Appendix~\hyperref[sec:AppA]{A}. More details on the method of the linear loss measurements are also given in ref.~\cite{espinosa2021tunable}.

\subsection{Nonlinear absorption}

To perform the nonlinear absorption analysis, it is convenient to calculate the light irradiance from the measured input and output powers. For the reference devices, the modal area does not change with the propagation distance. Therefore, the irradiance coupled into the waveguide, at its beginning [$I(0)$] and end [$I(L)$], is calculated by dividing the peak power by the effective mode area, according to
\begin{equation}
\label{eqn:I0}
    I(0) = \frac{0.94  (1-R) \eta_\text{C} P_\text{in}}{\tau_\text{FWHM} f_p A_\text{eff}}
\end{equation}
and
\begin{equation}
\label{eqn:IL}
    I(L) = \frac{0.94 P_\text{out}}{(1-R)\tau_\text{FWHM} f_p A_\text{eff}},
\end{equation}
respectively. Here $\eta_\text{C}$ is the coupling efficiency, and the factor $0.94$ comes from the assumption of a Gaussian temporal profile. For the tapered devices, on the other hand, the effective mode area changes along the propagation direction: it is the same as the reference devices for the coupling parts, and it is smaller for the nanosectional part. In order to evaluate the nonlinear absorption in the nanosectional part only, we performed the output vs. input power measurements as in the case of the reference devices, and then performed the analysis presented in subsection~\ref{sec:nano} to calculate the input and output irradiance data points at the nanosectional part only.

In a waveguide part with constant modal area, the change in the light irradiance $I(z)$ with the propagation distance $z$ is quantified by solving the differential equation
\begin{equation}
\label{eq:MPA}
    \frac{\text{d}I}{\text{d}z} = -\alpha_1 I - \alpha_2 I^2 - \alpha_\text{x} I^3,
\end{equation}
where the parameter $\alpha_2$ is the 2PA coefficient, and the parameter $\alpha_\text{x}$ can either be associated with the three-photon absorption (3PA) or FCA, as we will detail below. Here, the linear loss coefficient $\alpha_1$ quantifies the total effect of both the linear absorption and the light scattering. We note, however, that Eq.~(\ref{eq:MPA}) would be the same if $\alpha_1$ represented either the linear interband absorption alone or the light scattering alone. The reason is that $\alpha_1$ is just a measure of the exponential decrease of irradiance with propagation distance in the linear optical regime. For the nonlinear regime, we assume that the dominant irradiance-dependent loss mechanism is the nonlinear absorption.

When the nonlinear absorption is dominated by the 2PA mechanism ($\alpha_\text{x} = 0$), the solution of Eq.~(\ref{eq:MPA}) is given by %
\begin{equation}
\label{eq:2PA}
    I(z) = \frac{I(0)\text{exp}(-\alpha_1 z)}{1 + \alpha_2 I(0) L_\text{eff}},
\end{equation}
with $L_\text{eff}~=~[1 - \text{exp}(-\alpha_1 z)]/\alpha_1$. Conversely, if 2PA is negligible ($\alpha_2 = 0$) and the dominant mechanism is 3PA, $\alpha_\text{x}$ is the 3PA coefficient ($\alpha_\text{x} = \alpha_3$), and solution of Eq.~(\ref{eq:MPA}) is given by
\begin{equation}
\label{eq:3PA}
    I(z) = \frac{I(0)\text{exp}(-\alpha_1 z)}{\sqrt{1 + 2 \alpha_3 I^2(0) L'_\text{eff}}},
\end{equation}
where $L'_\text{eff}~=~ [1 - \text{exp}(-2 \alpha_1 z)]/(2 \alpha_1)$.

For the case where both $\alpha_2$ and $\alpha_x$ are non-vanishing, Eq.~(\ref{eq:MPA}) is usually solved numerically. One useful approximate solution is of the form~\cite{boudebs:2004}
\begin{equation}
\label{eq:MPAsol}
    I(z) = \frac{I(0)\text{exp}(-\alpha_1 z)}{1 +  \left({\alpha_2 - \frac{\alpha_1 \alpha_\text{x}}{\alpha_2}}\right) I(0) L_\text{eff} + \frac{\alpha_\text{x} \left[\alpha_1 + \alpha_2 I(0)\right]}{\alpha_2^2} \text{log}\left[ 1 + \alpha_2 I(0) L_\text{eff} \right]}.
\end{equation}
Eq.~(\ref{eq:MPAsol}) can be applied to the cases where both 2PA and 3PA are present. For semiconductors, the simultaneous absorption of three photons would promote an electron from the valence to the conduction band through two virtual levels. However, if the photon energy is enough to cause a 2PA process, the absorption of two photons already promotes the electron from the valence band to a real level at the conduction band, and the third photon would be absorbed by a free carrier. Therefore, for the wavelengths where 2PA occurs, we expect that 2PA followed by FCA would be more probable than a simultaneous 2PA and 3PA process.

For a mechanism where FCA follows 2PA, the coefficient $\alpha_\text{x}$ can be related to the FCA cross-section $\sigma$ and free-carrier density $N$ by~\cite{laughton1992intuitive}
\begin{equation}
\label{eq:alphax}
    \alpha_\text{x} = \frac{\sigma N}{I^2} = \frac{\sigma}{f(\tau)} \frac{ \alpha_2 \sqrt{\pi} T I^2}{4 \hbar \omega},
\end{equation}
where $T = \tau_\text{FHWM} / \sqrt{2 \text{ln} 2}$, and $f(\tau) = 1- \text{exp}(-t_{\text{p}}/\tau)$ is a function of the total free-carrier lifetime $\tau$, which accounts for the cumulative effect of the pulses~\cite{laughton1992intuitive}. The interval between pulses is $t_\text{p} = 1/f_\text{p}$. Note that the third term of the right-hand side of Eq.~(\ref{eq:MPA}), $\alpha_\text{x} I^3$, is equal to $\sigma N I$, as expected for a FCA process.

\section{Results}
\label{sec:res}

\subsection{Reference devices}
\begin{figure*}[h]
\centerline{\includegraphics[width=14cm]{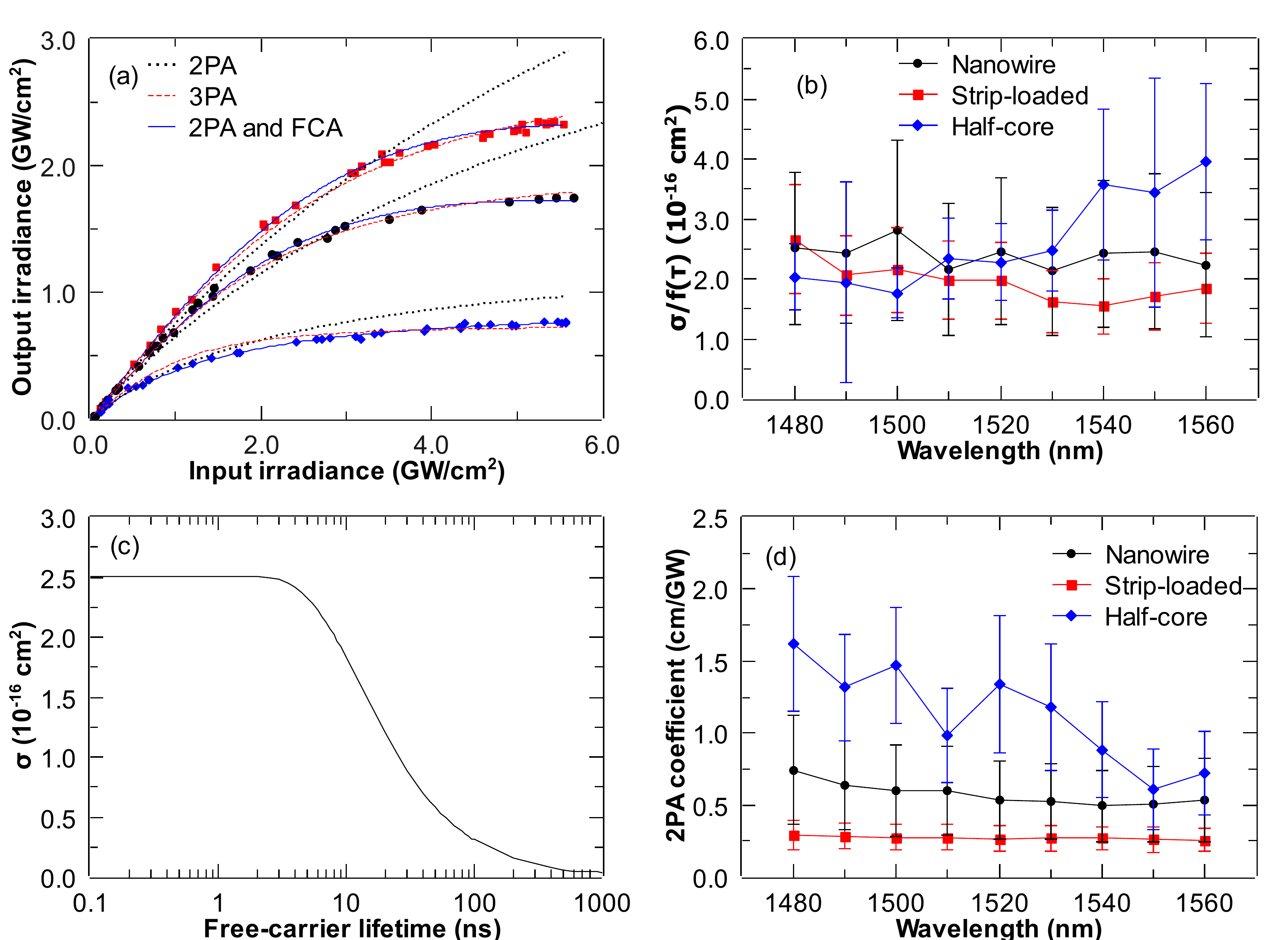}} \caption{\label{fig:reference}(color on-line) Results for the $2$-$\mu$m-wide reference devices. (\textcolor{red}{$\blacksquare$}) - strip-loaded, ($\bullet$) - nanowire, and (\textcolor{blue}{$\blacklozenge$}) - half-core. (a) Typical measurement of the output irradiance as a function of the input irradiance and the corresponding nonlinear fit at a wavelength of $1490~\text{nm}$. Legend: 2PA - fitting with Eq.~(\ref{eq:2PA}); 3PA - fitting with Eq.~(\ref{eq:3PA}); 2PA and FCA - fitting with Eq.~(\ref{eq:MPAsol}). (b) Free-carrier absorption cross-section normalized by $f(\tau)$, as a function of wavelength [$f(\tau) = 1 - \exp(-t_p / \tau)$]. (c) Free-carrier absorption cross-section calculated for different lifetimes for the $2$-$\mu$m-wide nanowire waveguide. (d) The experimental values of $\alpha_2$ as a function of wavelength.}
\end{figure*}

Fig.~\ref{fig:reference}~(a) presents typical plots of the output irradiance as a function of the input irradiance at $1490~\text{nm}$. The intensities are calculated from the average power obtained experimentally using Eqs.~(\ref{eqn:I0}) and (\ref{eqn:IL}). Fitting the data with Eq.~(\ref{eq:2PA}), with $I(0)$, $\alpha_1$ and $z = L$ as constants and $\alpha_2$ as a fitting parameter, does not result in good agreement, suggesting that another nonlinear absorption mechanism is present.
%

If 3PA alone is considered instead of 2PA, then Eq.~(\ref{eq:3PA}) should be used for the fit. Fitting the experimental results with Eq.~(\ref{eq:3PA}), with $I(0)$, $\alpha_1$, and $z = L$ as constants and $\alpha_3$ as a fitting parameter, results in a better outcome than fitting them with Eq.~(\ref{eq:2PA}). However, there are at least three indications against the interpretation of the nonlinear absorption as a 3PA mechanism. First, the 3PA coefficients $\alpha_3$ achieved by the fitting are $0.39$, $1.22$, and $1.49~\text{cm}^3\text{/GW}^2$ for the strip-loaded, half-core and nanowire geometries, respectively. These values are $10$ to $40$ times greater than the value measured in Ref.~\cite{aitchison:1997} at $1500~\text{nm}$. Second, from the 3PA scaling rule, it is expected that the 3PA coefficient increases with wavelength~\cite{aitchison:1997}, but fitting the data at other wavelengths with Eq.~(\ref{eq:3PA}) results in $\alpha_3$ decreasing as $\lambda$ increases. Third, the bandgap of the guiding layer material (Al$_{0.18}$Ga$_{0.82}$As) is about $(1.66\pm0.02)~\text{eV}$. This value was calculated by taking the mean and standard deviation of the bandgap energy reported in Refs.~\cite{el1993experimental,huang1988photoreflectance,bosio1988direct,wrobel1987variations,oelgart1987photoluminescence,aspnes1986optical,aubel1985interband,miller1985accurate,lee1980electron,casey1978room,monemar1976some}. From the bandgap value, the two-photon absorption edge is calculated as $(1494\pm18)~\text{nm}$, and the three-photon absorption edge as $(2241\pm27)~\text{nm}$. Therefore, at a wavelength of $1490~\text{nm}$, a 3PA transition is not expected. The energy of two photons is enough to overcome the bandgap and promote a 2PA transition from the valence to the conduction band. The third photon of a hypothetical 3PA transition, in this case, is absorbed by an electron promoted to the conduction band, i.e., by a free carrier. The overall process then is 2PA followed by FCA, not a 3PA process. Thus, the fitted $\alpha_3$ is an effective coefficient arising from the 2PA and FCA transitions instead of the actual 3PA process, which explains such a high $\alpha_3$ value.

In this case, considering the mechanism of a 2PA transition followed by a free-carrier absorption would be more appropriate than extracting a 3PA coefficient. The 2PA-FCA analysis comprises fitting the data with Eq.~(\ref{eq:MPAsol}). The fitting parameters are $\alpha_x$ and $\alpha_2$, while the constants are $z = L$, $\alpha_1$, and $I(0)$. The agreement between the fitting curve and experimental results in this case is even better than that obtained from fitting the results with Eq.~(\ref{eq:3PA}). For the reference devices, the normalized FCA cross-section, calculated by using the extracted fit parameter $\alpha_x$ and Eq.~(\ref{eq:alphax}) are presented in Fig.~\ref{fig:reference}~(b) while the $\alpha_2$ results are presented in Fig.~\ref{fig:reference}~(d). Note that one could naively consider $\alpha_x$ as the 3PA coefficient ($\alpha_3$) in an eventual simultaneous 2PA and 3PA effect. However, in this case, the same arguments against this interpretation given in the previous paragraph apply: $\alpha_3$ results are one order of magnitude higher than what is presented in Ref.~\cite{aitchison:1997}, $\alpha_3$ decreases for longer wavelengths, and the bandgap value allows 2PA followed by FCA.


The actual FCA cross-section of each waveguide can be achieved by multiplying $\frac{\sigma}{f(\tau)}$, presented in Fig.~\ref{fig:reference}~(b), by $f(\tau)$. Still, one needs to know the free-carrier lifetime in order to calculate $f(\tau)$. If we assume $\tau = 5~\text{ns}$, the same value as that used in Ref.~\cite{villeneuve1994nonlinear} for a GaAs/AlGaAs waveguide, then $f(\tau) = 0.927$ and the resulting FCA cross-sections are very close to what was reported in that reference. However, the lifetime found in the literature for AlGaAs waveguides, GaAs/AlGaAs waveguides, and GaAs/AlGaAs quantum well structures, ranges from $0.06~\text{ns}$ to $35~\text{ns}$~\cite{szymanski2009ultrafast, ma2014two,apiratikul2009semiconductor,orton1994recombination,laughton1992time}. To evaluate what the FCA cross-section would be if a different $\tau$ was adopted, we used Eq.~(\ref{eq:alphax}) and calculated $\sigma$ for a wide range of lifetimes. The result is presented in Fig.~\ref{fig:reference}~(c). We note that for $\tau \lessapprox 35~\text{ns}$ [$f(\tau) \gtrapprox 0.3$], the order of magnitude of $\sigma$ would still be $10^{-16}~\text{cm}^2$, in agreement with the FCA cross-section recently obtained in the 2PA and 3PA spectral region of bulk GaAs~\cite{benis2020three}.

Let us discuss further in the results presented in Fig.~\ref{fig:reference}~(b). Assuming the same lifetime for all geometries, the FCA cross-section of the strip-loaded and nanowire geometries slightly decreases when the wavelength increases. For the half-core device, $\sigma$ is constant, given the error, but increases for longer wavelengths, which possibly indicates some residual effect of 3PA in this region~\cite{benis2020three}.

The 2PA coefficients [Fig.~\ref{fig:reference}~(d)] decrease with increasing wavelength for all three geometries. We expect this behavior based on the 2PA scaling law, also considering that we are working near half-the-bandgap region~\cite{wherrett:1984}. However, $\alpha_2$ does not drop to zero above half-the-bandgap [wavelengths longer than $(1494\pm18)~\text{nm}$] but continues to exponentially decrease, exhibiting a typical band tailing. Other works have reported an Urbach tail type of behavior for the two-photon absorption in AlGaAs, originating from defect states and impurities~\cite{stegeman1994algaas,kang1994limitation,villeneuve:1993}. Yet, the exact form and limits of the band tail were not determined in this work due to the limitation of the illuminating wavelength range of our experimental setup.

\subsection{Nanosectional parts}
\label{sec:nano}

\begin{figure}[h]
\centerline{\includegraphics[width=8cm]{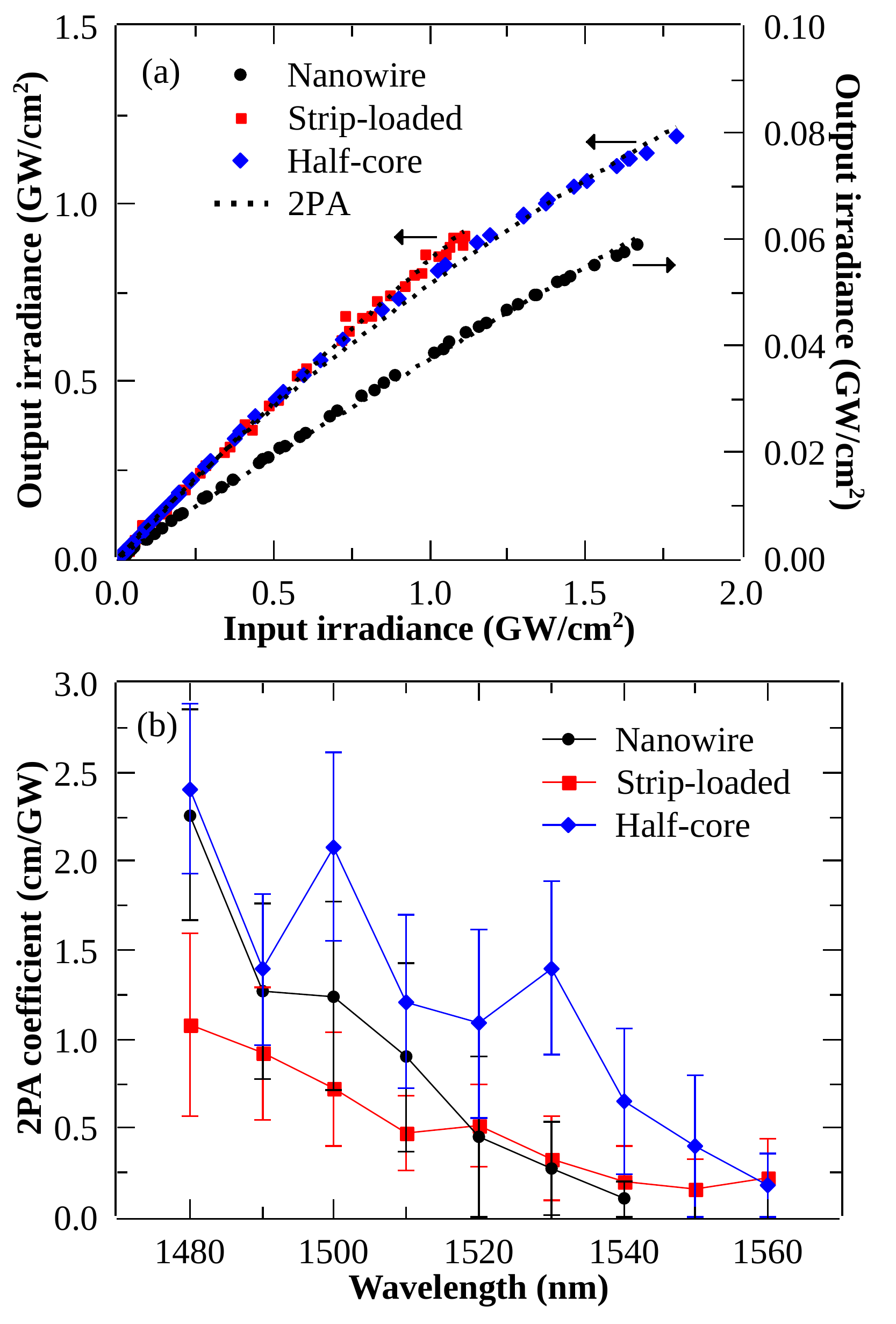}} \caption{\label{fig:nanosectional}(color on-line) Results for the nanosectional parts of the tapered devices. (\textcolor{red}{$\blacksquare$}) - strip-loaded, ($\bullet$) - nanowire, and (\textcolor{blue}{$\blacklozenge$}) - half-core. (a) Typical measurement of the output irradiance as a function of the input irradiance and the corresponding nonlinear fit with Eq.~(\ref{eq:2PA}) at a wavelength of $1490~\text{nm}$. (b) The experimental values of $\alpha_2$ as a function of wavelength.}
\end{figure}

The input ($P_\text{in}$) and output ($P_\text{out}$) power were also measured for the tapered devices by using the nonlinear transmittance technique. However, to characterize the nonlinear absorption at the nanosectional part only, we need to calculate the irradiance at its beginning ($z = L_\text{c} + L_\text{t}$) and end ($z = L_\text{c} + L_\text{t} + L_\text{NS}$) [see Fig.~\ref{fig:waveguide-designs}~(a) to locate the $z$ positions presented in this analysis]. Hence, the losses in the coupling part have to be subtracted from the overall loss in order to focus on the loss experienced at the nanosectional part only. In the following discussion, we present a method in $6$ steps to perform the calculation of the irradiance at the nanosectional part of the devices. Then, the results achieved with this method are presented in Fig.~\ref{fig:nanosectional}.

\begin{enumerate}[]

\item \textbf{Determining the irradiance at the positions $z = 0$ and $z = L$.} The irradiance at the input [$I(0)$] and output [$I(L)$] of the tapered device were calculated with Eq.~(\ref{eqn:I0})~and~(\ref{eqn:IL}), respectively, by using the $P_\text{in}$ and $P_\text{out}$ experimental data.

\item \textbf{Determining the irradiance at the position $z = L_\text{C}$.} We suppose that a 2PA process followed by FCA also occurs at the input coupler (the first part of the waveguide with length $L_\text{C}$). Then, the irradiance at the end of the input coupling part [$I(L_\text{C})$] was calculated by Eq.~(\ref{eq:MPAsol}). The other parameters used in this calculation, besides $z = L_\text{C}$ and $I(0)$, are $\alpha_1$, $\alpha_2$ and $\alpha_x$ of the reference devices.

\item \textbf{Determining the irradiance at the position $z = L_\text{C} + L_\text{t}$.} The irradiance at the input of the nanosectional part [$I(L_\text{C} + L_\text{t})$] was calculated from
\begin{equation*}
    I(L_\text{C} + L_\text{t}) = t_{\text{l}} \frac{ A_\text{eff}^{\text{C}}}{ A_\text{eff}^{\text{NS}}} I(L_\text{C}),
\end{equation*}
where $A_\text{eff}^{\text{C}}$ and $A_\text{eff}^{\text{NS}}$ are the effective area of the coupling and nanosectional parts, respectively. The linear loss in the taper parts is given by the taper loss coefficient ($t_\text{l}$), which is the fraction of power transmitted through each $0.2$-mm-long taper. It was determined as $t_\text{l} = 0.89$ for the nanowire and $t_\text{l} = 1$ for the strip-loaded and half-core geometries, by comparing the transmission of reference and taper-to-taper ($L_{\text{NS}} = 0$) devices. Due to the short taper length in comparison to $L_\text{C}$ or $L_\text{NS}$, the nonlinear absorption in the tapers was neglected.

\item \textbf{Determining the irradiance at the position $z = L - L_\text{C}$.} To calculate $I(L - L_\text{C})$, we supposed that there is only linear loss at the output coupler, so
\begin{equation*}
    I(L - L_\text{C}) = e^{\alpha_1 L_\text{C}} I(L) ,
\end{equation*}
with $\alpha_1$ of the reference devices, and $I(L)$ calculated in step $1$.

\item \textbf{Determining the irradiance at the position $z = L_\text{C} + L_\text{t} + L_\text{NS}$.} The irradiance at the output of the nanosectional part [$I(L_\text{C} + L_\text{t} + L_\text{NS})$] was calculated from
\begin{equation*}
I(L_\text{C} + L_\text{t} + L_\text{NS}) =  \frac{A_\text{eff}^{\text{C}}}{t_{\text{l}} A_\text{eff}^{\text{NS}}} I(L - L_\text{C}). 
\end{equation*}

\item \textbf{Plotting the data points}. The x- and y-coordinates of the data points presented in Fig.~\ref{fig:nanosectional}~(a) were then taken as $I(L_\text{C} + L_\text{t})$ (from step $3$) and $I(L_\text{C} + L_\text{t} + L_\text{NS})$ (from step $5$), respectively.

\end{enumerate}

Fig.~\ref{fig:nanosectional}~(a) presents the output irradiance [$I(L_\text{C} + L_\text{t} + L_\text{NS})$] as a function of the input irradiance [$I(L_\text{C} + L_\text{t})$] at the nanosectional parts. Here, Eq~(\ref{eq:2PA}) fits the data well, indicating the absence of FCA effects. The linear loss and nonlinear absorption cause the irradiance to drop along the coupling part such that the irradiance in the nanosectional part is sufficient to measure the 2PA coefficients, but too low to observe the effect of FCA. The corresponding $\alpha_2$ values are presented in Fig.~\ref{fig:nanosectional}~(b). As in the case of the reference devices, the 2PA coefficients also decrease with increasing wavelength for all three geometries, in accordance with the 2PA scaling law~\cite{wherrett:1984}. We also observe the Urbach-like band tailing in the nanosectional parts.

\section{Discussion}
\label{sec:dis}

Let us compare the $\alpha_2$ results with the previous work from literature. For the strip-loaded geometry with Al$_{0.18}$Ga$_{0.82}$As as the guiding layer, the earlier reported values at the wavelength of $1500~\text{nm}$ (TE mode) are $\alpha_{2} = (1.0 \pm 0.3)$, $0.8$, and $(1.15 \pm 0.15)~\text{cm/GW}$, presented in Refs.~\cite{aitchison:1997},~\cite{kang1994limitation}, and~\cite{villeneuve:1993}, respectively. These values agree well with our results for the nanosectional part of the strip-loaded geometry: $(0.7 \pm 0.3)~\text{cm/GW}$. The spectral behaviour is also very similar to that presented in Fig.~\ref{fig:nanosectional}~(b). There is no reported $\alpha_2$ for the waveguides of nanowire geometry whose guiding layer is made of Al$_{0.18}$Ga$_{0.82}$As, although Ref.~\cite{wathen2014efficient} reports $\alpha_{2} = 3.3~\text{cm/GW}$ for a $1.5$-$\mu$m-wide nanowire waveguide, with Al$_{0.13}$Ga$_{0.87}$As as the guiding layer, at $1550~\text{nm}$. Moreover, the present work is the first to report on the 2PA coefficient measurement of the half-core waveguide geometry.

For both the reference and tapered devices, the value of $\alpha_2$ is the highest for the half-core design, followed by the nanowire and then the strip-loaded. The difference between the results obtained for the three geometries is thought to be caused by the different nonlinear confinement factors or defects due to fabrication imperfection, as discussed further.

To further compare the different geometries, let us now discuss the confinement factor. The modal field is not fully confined in the guiding layer region of the waveguides. It spreads into the claddings and the surrounding media in a geometry-dependent way [see the field distribution in Fig.~\ref{fig:waveguide-designs}~(c)-(h)]. Therefore, what Fig.~\ref{fig:reference}~(d) and Fig.~\ref{fig:nanosectional}~(b) present are the effective 2PA coefficients. Here, effective means the average value weighted by the irradiance of each material. 

In Ref.~\cite{grant1996effective}, the authors proposed the formula to calculate the effective nonlinear refractive index and indicated how to adapt their approach to calculate the effective 2PA coefficient ($\alpha_{2,\text{eff}}$). Based on their suggestion, we derived
\begin{align}
\alpha_{2,\text{eff}} & = \Gamma_{2,(1)} \alpha_{2,(1)} + \Gamma_{2,(2)} \alpha_{2,(2)} + ...,\label{eq:alphaeff}
\end{align}
where $\Gamma_{2,M}$ ($\alpha_{2,M}$) is the third-order nonlinear confinement factor (2PA coefficient) for each material (or each layer), $M$, of the waveguide structure. The third-order nonlinear confinement factor is calculated by
\begin{align}
\Gamma_{2, M} & = \frac{ \int_{M} I^2(x,y) \text{d}A}{ \int_{-\infty}^\infty I^2(x,y) \text{d}A}.\label{eq:confinament}
\end{align}
The integral in the numerator of Eq.~(\ref{eq:confinament}) is performed only over the area of the material $M$ (either the guiding layer or the cladding), and $I(x,y)$ denotes the irradiance at the position $(x,y)$ of the transverse mode~\cite{grant1996effective}. The same approach can be used to express the effective 3PA coefficient
\begin{align}
\alpha_{3,\text{eff}} & = \Gamma_{3,(1)} \alpha_{3,(1)} + \Gamma_{3,(2)} \alpha_{3,(2)} + ...,\label{eq:alphaeff2}
\end{align}
but in this case, the weighting factor is the fifth-order confinement factor given by~\cite{grant1996effective}
\begin{align}
\Gamma_{3, M} & = \frac{ \int_{M} I^3(x,y) \text{d}A}{ \int_{-\infty}^\infty I^3(x,y) \text{d}A}.\label{eq:confinament2}
\end{align}
$\Gamma_{2, M}$ and $\Gamma_{3, M}$ can be calculated with FDE solver of \textit{Lumerical Mode Solutions}, using the \textit{integrate} function to implement Eq.~(\ref{eq:confinament}) and Eq.~(\ref{eq:confinament2}), respectively.

As we discussed earlier, we have dismissed 3PA in Al$_{0.18}$Ga$_{0.82}$As (guiding layer) because the energy of two photons is greater than the bandgap so 2PA (and FCA) are the dominant mechanisms.

Let us now discuss the bandgap energy and nonlinear absorption mechanisms of the cladding materials [see fig.~\ref{fig:waveguide-designs}~(c)-(h) for the material composition of each layer]. Al$_{0.35}$Ga$_{0.65}$As has a bandgap energy of approximately $1.93~\text{eV}$~\cite{bosio1988direct}. Thus, 2PA is not expected for $\lambda \gtrsim 1288~\text{nm}$ but 3PA can occur for $\lambda \lesssim 1927~\text{nm}$. Similarly, the bandgap of Al$_{0.65}$Ga$_{0.35}$As is $2.40~\text{eV}$~\cite{bosio1988direct}, making 2PA negligible for $\lambda \gtrsim 1033~\text{nm}$ but allowing 3PA for $\lambda \lesssim 1550~\text{nm}$. Therefore, it is reasonable to assume $\alpha_{2,\text{cladd}} \Gamma_{2, \text{cladd}} = 0$ for either Al$_{0.35}$Ga$_{0.65}$As and Al$_{0.65}$Ga$_{0.35}$As so that the effective 2PA coefficient is the contribution of the guiding layer only, i.e.,
\begin{align}
\alpha_{2,\text{eff}} = \Gamma_{2,\text{guid}} \alpha_{2,\text{guid}}.\label{eq:confcore}
\end{align}

Conversely, the contribution of 3PA in the claddings is calculated from $\alpha_{3,\text{cladd}} \Gamma_{3, \text{cladd}}$ for each cladding material. However, $\Gamma_{3,\text{cladd}} < 0.03$ for the Al$_{0.65}$Ga$_{0.35}$As layers in all waveguides geometries, and the effect of 3PA in the overall absorption due to this material can be neglected at the irradiance levels of this work. Al$_{0.35}$Ga$_{0.65}$As, on the other hand, is only present in the composition of the strip-loaded waveguides. $\Gamma_{3} = 0.23$ and $0.08$ for the upper cladding of the $2$-$\mu$m-wide and $0.9$-$\mu$m-wide strip-loaded waveguides, respectively. Since the strip-loaded has the lowest nonlinear absorption of all geometries, we did not include the eventual effect of 3PA in the Al$_{0.35}$Ga$_{0.65}$As upper cladding when extracting the 2PA coefficients.

\begin{figure}[h]
\centerline{\includegraphics[width=8cm]{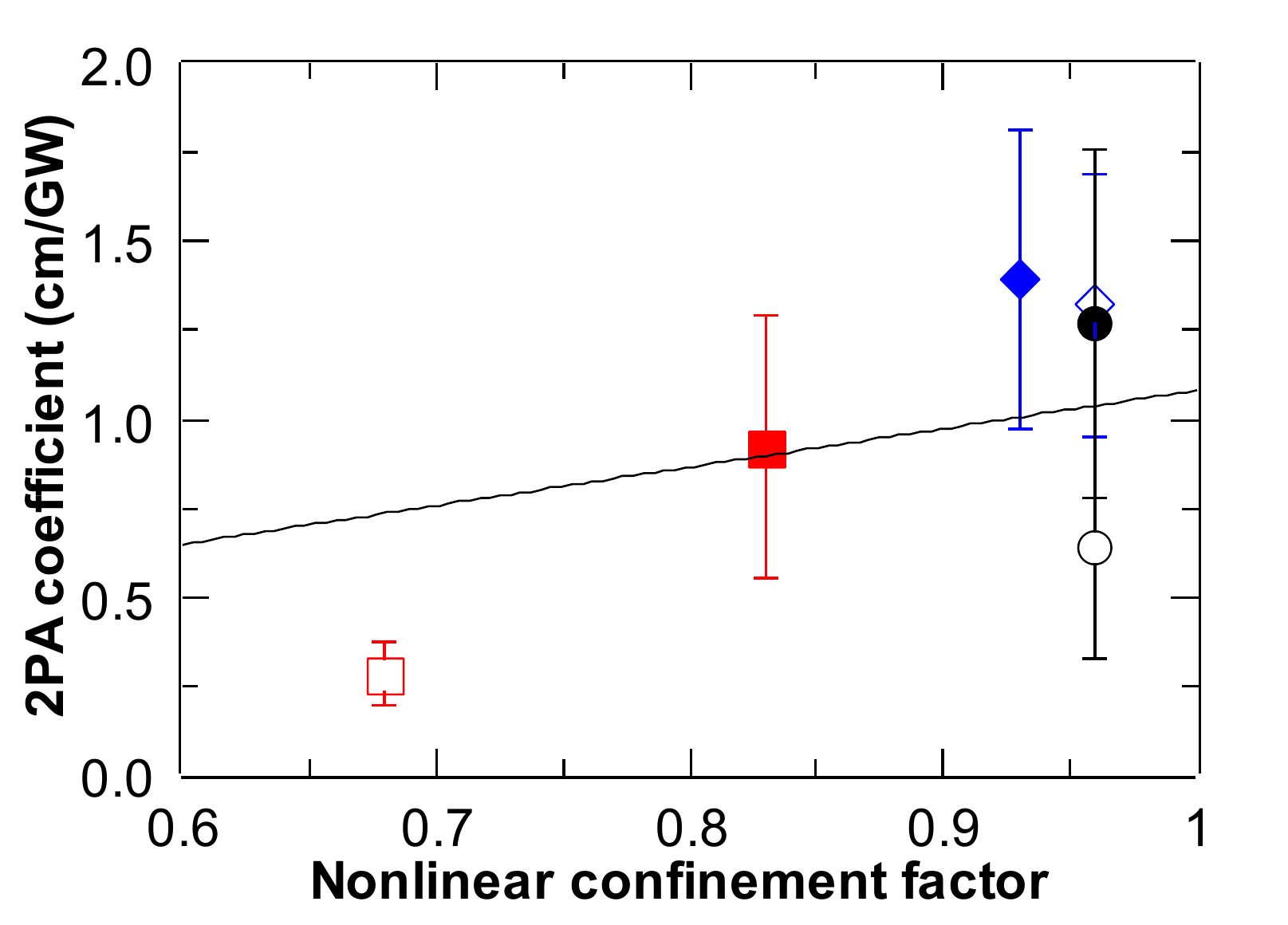}} \caption{\label{fig:CF}(color on-line) Effective $\alpha_2$ at $1490~\text{nm}$ as a function of the waveguide's guiding layer nonlinear confinement factor $\Gamma_{2,\text{guid}}$.  $2$-$\mu$m-wide waveguides: (\textcolor{red}{$\square$}) - strip-loaded, ($\circ$) - nanowire, and (\textcolor{blue}{$\lozenge$}) - half-core. Nanosectional parts: (\textcolor{red}{$\blacksquare$}) - strip-loaded, ($\bullet$) - nanowire, and (\textcolor{blue}{$\blacklozenge$}) - half-core. The solid line is a fitting with Eq.~(\ref{eq:confcore}), resulting in $\alpha_{2,\text{guid}} = 1.1~\text{cm}/\text{GW}$.}
\end{figure}

Let us now suppose that the guiding layer of the three geometries is made solely from Al$_{0.18}$Ga$_{0.82}$As. Let us also assume that the defects and impurities concentration in the guiding layer are the same for the three geometries. Then the material $\alpha_{2,\text{guid}}$ should be the same for the three geometries. Based on Eq.~(\ref{eq:confcore}), the effective 2PA coefficient should be different for each geometry or each waveguide part if the nonlinear confinement factors are not the same. Fig.~\ref{fig:CF} presents the effective 2PA coefficient as a function of the nonlinear confinement factor of the waveguide's guiding layer for the three geometries at the nanosectional part and for the reference device. $\Gamma_{2,\text{guid}}$ was calculated with FDE solver of \textit{Lumerical Mode Solutions}, using the \textit{integrate} function to implement Eq.~(\ref{eq:confinament}). The effective 2PA coefficient slowly increases with the increase of the nonlinear confinement factor. By fitting the data with Eq.~(\ref{eq:confcore}), we found $\alpha_{2,\text{guid}} = 1.1~\text{cm}/\text{GW}$. However, some points do not agree with the fitted curve, which is an indication that other factors, besides $\Gamma_{2,\text{guid}}$, influence our $\alpha_2$ results.

Fabrication imperfections might have affected our results in two ways. First, the sidewall roughness of the waveguide's guiding layer can be present in the nanowire and half-core geometry because the guiding layer (or part of it) is etched through [Fig.~\ref{fig:waveguide-designs}~(g)~and~(h)]. The strip-loaded geometry, on the other hand, is less susceptible to such roughness because the guiding layer is fully covered by the upper cladding [Fig.~\ref{fig:waveguide-designs}~(f)]. Hence, nonlinear scattering or defect states might have contributed to the higher values of the 2PA coefficient for the nanowire and half-core geometries. Second, we used the characteristics of the reference waveguides [Fig.~\ref{fig:waveguide-designs}~(b)] to evaluate the linear and nonlinear absorption contributions of the coupling parts of the device [see Fig.~\ref{fig:waveguide-designs}~(a)]. However, since the reference waveguide is a separate device, its characteristics might slightly differ from those of the coupling parts, thereby affecting the precision of determining the nanosectional part's 2PA coefficient.

Based on our results, we highlight two important considerations for the evaluation of nonlinear absorption in waveguides. First, the effective 2PA coefficient is geometry-dependent and may be affected by fabrication imperfections. For this reason, the measurement of 2PA coefficient in the actual fabricated device, rather than taking the $\alpha_2$ value from the literature for similar material composition, is preferable. Second, Eqs.~(\ref{eq:alphaeff}) and (\ref{eq:confinament}) can be used to evaluate the effective 2PA for specific compositions and geometries when designing the waveguides. The aluminum fraction $x$ is often used as a design parameter to determine the 2PA edge wavelength and how strong 2PA would be at a particular wavelength range~\cite{stegeman1994algaas}. Our work suggests that $\Gamma_{2,\text{guid}}$ can also be used as a design parameter to evaluate the geometry-dependent 2PA.

\section{Conclusion}
\label{sec:conc}

We measured the 2PA coefficients of AlGaAs waveguides of three different geometries. The free-carrier absorption effect was also included in our analysis for an accurate interpretation of the experimental data. We noticed that although these devices have the same composition in their guiding layers (Al$_{0.18}$Ga$_{0.82}$As), they offer different 2PA coefficient values. We explained this observation by including the effect of a geometry-dependent electrical field confinement in the guiding region of the devices. As the three AlGaAs layers comprising the waveguides have different aluminum fractions, the measured effective 2PA coefficient depends on whether the irradiance is more or less confined in one or another layer. We used the third-order nonlinear confinement factor to evaluate this effect. On average, the half-core geometry presented higher 2PA coefficients, followed by the nanowire and then the strip-loaded waveguide. Therefore, measuring the 2PA coefficient for a new waveguide geometry and/or material composition is better than taking the corresponding values from the literature.

\appendix

\section*{Appendix A: Linear loss}
\label{sec:AppA}
%
\begin{figure}[h]
\centerline{\includegraphics[width=8cm]{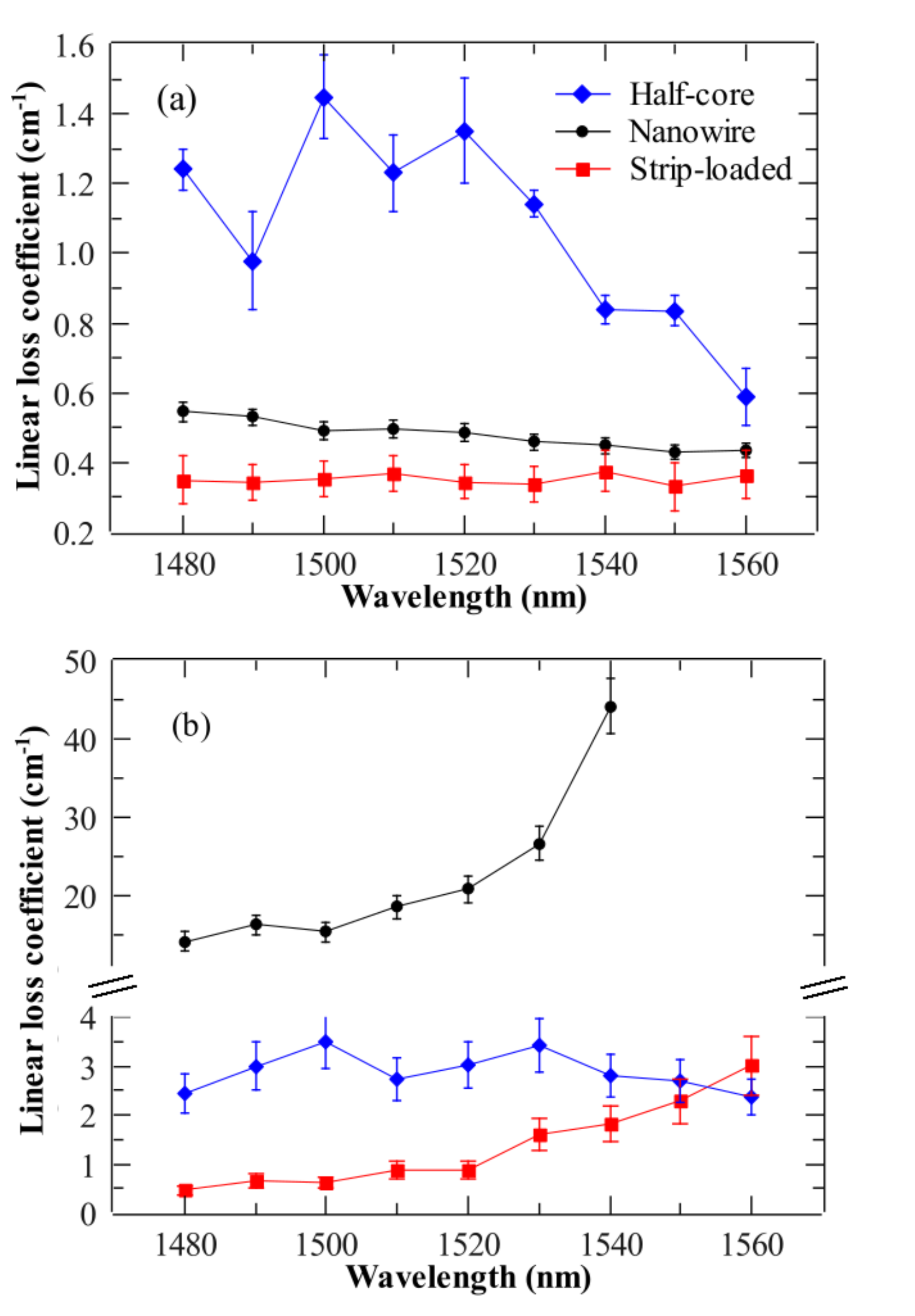}} \caption{\label{fig:alpha1}(color on-line) Linear loss coefficient $\alpha_1$ as a function of wavelength for the $2$-$\mu$m-wide waveguides (a) and for the nanosectional parts (b).}
\end{figure}

The linear loss coefficients of the reference devices and the nanosectional parts are presented in Fig.~\ref{fig:alpha1}~(a)~and~(b), respectively. As the wavelength range of this work falls in the transparency window of the waveguide materials, the light scattering is the major contribution to the linear loss, besides some residual absorption from impurities of defects in the materials. For the strip-loaded and nanowire waveguides, the linear loss is almost constant across the whole wavelength range. Yet, it is higher in the nanowires than in the strip-loaded waveguides. This is thought to be caused by the sidewall roughness of the exposed sidewalls of the nanowire guiding layer, leading to higher scattering losses in the nanowires~\cite{dolgaleva:2015}. Since the mode in the nanowire is confined in the region that is etched through, the sidewall surface exposed to air tends to have surface roughness. In addition, any oxidation on the sidewall's surface can impact the mode's propagation. In contrast, the guiding layer region in the strip-loaded waveguides is buried underneath the upper cladding. The interfaces between AlGaAs layers tend to be smoother than the etched sidewalls, leading to fewer scattering centers. We expect the linear loss coefficient of the half-core geometry to be in between that of the strip-loaded and nanowire since only a part of the guiding layer region is exposed to the air with sidewall roughness. This trend is observed for the narrowest parts of the waveguides [Fig.~\ref{fig:alpha1}~(b)]. However, for the reference devices, the half-core has the highest liner loss, which fluctuates from wavelength to wavelength [Fig.~\ref{fig:alpha1}~(a)], possibly because of a higher scattering at the waveguide's input and output facets~\cite{espinosa2021tunable}. As a result, the uncertainty in determining the 2PA coefficient and FCA cross-section is higher for the half-core geometry.

\begin{backmatter}
\bmsection{Funding}
The financial support was provided by Canada First Research Excellence Fund ``Transformative Quantum Technologies'', the Natural Sciences and Engineering Council of Canada (Discovery and RTI programs), the Canada Research Chairs program, the Canadian Foundation of Innovations, and the Ontario Early Researcher Award. The epitaxial growth of AlGaAs wafers was supported by a CMC Microsystems award.


\bmsection{Disclosures}
The authors declare no conflicts of interest








\bmsection{Data availability} Data underlying the results presented in this paper are not publicly available at this time but may be obtained from the authors upon reasonable request. 




\end{backmatter}


\bibliography{2PAFCA}

\end{document}